\documentclass[preprint,eqsecnum,aps,showkeys,showpacs]{revtex4}
\usepackage{dcolumn}
\usepackage{graphicx}
\usepackage{amsmath}
\usepackage{amssymb}
\usepackage{epsfig}
\usepackage{float}
\usepackage{latexsym}
\usepackage{rotating}

\begin{document}

\title{ \small Black Hole Solutions and Pressure Terms in Induced Gravity with Higgs Potential}

\vspace{4.0cm}
\author{Hemwati Nandan$^{1,3}${\footnote{hnandan.ctp@jmi.ac.in}}, Nils M. Bezares-Roder$^{2,3}${\footnote{Nils.Bezares@gmail.com}} and Heinz Dehnen$^3${\footnote{Heinz.Dehnen@uni-konstanz.de}}}
\vspace{0.2cm}
\affiliation{ \small $^1$Centre for Theoretical Physics, Jamia Millia
  Islamia, $110 \, 025$  New Delhi, India\\
$^2$Institut f\"ur Theoretische Physik, Universit\"at Ulm,$\,890 \,69$ Ulm, Germany\\
$^3$Fachbereich Physik, Universit\"at Konstanz, $M 677,
784 \, 57$ Konstanz, Germany}
\begin{abstract}
We study the quintessential properties of the Black Hole solutions in a scalar--tensor theory of gravity with Higgs potential in view of the static and spherically symmetric line element. The Reissner--Nordstr\"om-like and Schwarzschild Black Hole solutions are derived with the introduction of a series-expansion method to solve the field equations with and without Higgs field mass. The physical consequences of the Black Hole solutions and the solutions obtained in the weak field limit are discussed in detail by the virtue of the equation-of-state parameter, the scalar-field excitations and the geodesic motion. The appearance of naked singularities is also discussed together with the dependence of Black Hole horizons on the field excitations, which are themselves dependent on pressure terms which effectively screen the mass terms. A possible connection to flat rotation curves following the interaction with the scalar field is also presented in the weak field limit of gravity, together with a discussion of dynamical effects of scalar fields and pressure terms on mass.
\end{abstract}
\pacs{97.60.Lf, 14.80.-j, 95.35.+d, 95.36.+x, 11.15.Ex 11.30.-j, 97.60.Lf.}

\vspace{0.5cm}
\keywords{Symmetry Breaking, Higgs field, Scalar--Tensor Theory,
Singularity, Black Holes, Quintessence and Dark Matter.}

\maketitle

\section{Introduction}
\noindent  Scalar fields in different models of scalar--tensor
theories of gravity have played crucial role in addressing various issues in gravity and cosmology in diverse contexts (viz problem of dark energy, inflation, late time acceleration of the Universe, Quintessence, cosmic evolution etc.) in recent years \cite{Copeland}. Besides these issues, the appearance of Black Hole (BH), singular and nonsingular solutions in a scalar--tensor theory \cite{Campanelli,Bezares07-BH}  with some other closely related issues \cite{Avelino} are also of crucial interest. In particular, the naked singularities asserting the violation of the weak cosmic censorship hypothesis (WCCH) \cite{Penrose98} seem to appear in many realisations of gravitational collapse for spherical \cite{Ori} and non-spherical symmetric systems \cite{Shapiro91}. At observational front, such violation of WCCH would inherently be connected to some sorts of gamma-ray bursts (GRBs) \cite{Chakrabarti} and gravitational lensing by naked singularities \cite{ksv}.\\
The use of a Higgs-like scalar field in a Jordan--Brans--Dicke (JBD) model of gravity \cite{Dehnen1, Dehnen2, Dehnen3, Cervantes95a, Bezares07-DM} is of quite significant importance due to the role of Higgs potentials in mass generation mechanisms and as cosmological-function (anti-gravitational) term and Quintessence \cite{Cervantes07a}. In fact, anti-gravitational behaviour in such model leads to Schwarzschild BHs which do not appear in the usual sense for non-vanishing and decaying scalar-field excitations. Both the metric and scalar field are regular everywhere for non-vanishing excitations with exception of $r=0$ as naked singularity \cite{Bezares07-BH}. For negative scalar-field excitations, the metric components evolve with three different patterns depending on the amplitude of the scalar-field excitation where the horizons vanish for the central symmetry. It is therefore meaningful to further investigate the nature of quintessential behaviour of the BH and singular solutions of the field equations of this model with the precise role of scalar-field excitations ($\xi$) in analogy to the well known Reissner--Nordstr\"{o}m (RN) solutions \cite{Rei16}. In order to relate new effects to usual dynamical quantities, we have therefore addressed the quintessential properties of the present induced model of gravity in the context of RN solutions which are valid especially for non-strong gravitational regimes. Such  analysis also portrays the role of scalar field and pressure in the dynamics in view of the vanishing of singularities and BH solutions.\\
The plan of this paper is as follows: we first briefly recall the formalism of the scalar--tensor theory of gravity with a Higgs field in Section II. We look for the solutions of the field equations for non-vanishing scalar-field excitations by using a series-expansion method in Section III. A RN-like BH solution having anti-gravitational features is derived, and both the mass and charge parameters of the solution are found to depend on the pressure terms related to the scalar field. For physical interpretation of the solutions obtained which are also valid for weak fields,  we compare them with the linear solutions of an almost-flat metric in Section IV. The linear Poisson equation comprising the scalar field possesses pressure terms, and these act within dynamical, effective masses. In view of the geodesic motion, we then implement the results of Section III on rotation curves in Section V and discuss the effective potential in terms of its bounding properties in view of the large astronomical distances and relatively weak gravitational fields as expected beyond galactic bulges. We also discuss the notion of effective and dynamical mass in contraposition to the luminous mass as derived from usual density. Finally, we conclude our findings with possible future directions in Section VI.

\section{Motivation and Basic Formalism:  A Brief Overview}

\noindent Let us quickly review the formalism of the scalar--tensor theory of gravity with a Higgs potential in this Section. Remarkably, both the Brans--Dicke and Higgs scalar fields are inherently well-connected to the problem of \emph{origin of mass} in nature, which is evident from their following role respectively:\\
\noindent (i) In a JBD theory, a scalar field shares the stage with
gravity, and as such it is compatible with the notions of Mach's principle (i.e. inertia as gravitation) \cite{Fujii}. The gravitational constant thus depends on the (gravitational) mass distribution in the Universe, and it is replaced with the inverse of the Brans--Dicke scalar field. Further, in a prototype JBD model with a broken-symmetric theory of gravity motivated from the smallness of the coupling constants of weak and gravitational interactions as well as the commonality in their dimensions, the gravitational coupling itself can be generated in terms of the VEV of a scalar field \cite{Zee}.\\
\noindent (ii) A Higgs field is source of the inertial mass of the elementary particles in the standard model (SM) and various other unification schemes. The elementary constituents (particles) in such theories acquire mass only in the broken phase of symmetry while they are massless before the breakdown of the symmetry. A Higgs field may therefore serve as a scalar field in a scalar--tensor theory of gravity in view of the equivalence principle in GR \cite{Dehnen3}. The gravitational constant in the broken phase of symmetry of such theory can be then recast in terms of the VEV of a Higgs field with the consideration of following action \cite{Dehnen1, Dehnen2, Dehnen3},
\begin{equation}
{\cal S}= \int d^4 x\, \sqrt{-g}\, \left[\frac{\gamma}{16\pi}\,
\phi^\dagger \phi \, {R} \,+ \, \frac{1}{2} \, \nabla _\mu \phi^\dagger \,
\nabla^\mu \phi - V(\phi) + {\cal L}_M (\psi, A_\mu, \phi)
\right], \label{action}
\end{equation}
where $\gamma$ is a dimensionless constant and ${\cal L}_M (\psi, A_\mu, \phi)$ is the Lagrangian corresponding to matter (i.e. it contains the fermionic $(\psi)$, massless bosonic ($A_\mu$) and scalar ($\phi$) fields). The self-interacting Higgs potential $V(\phi)$ in (\ref{action}) characterises the varying cosmological and gravitational constants. The potential is normalised such that it realise Zee's assumption $V(\phi = v) =0$ for the ground-state value of $\phi$ \cite{Zee} with the following form,
\begin{equation}
V(\phi)= \frac{\mu^2}{2} \, \phi \, \phi^\dagger + \frac{\lambda}{4!}
\, (\phi^\dagger \phi)^2+ \bar{V}, \label{H}
\end{equation}
where $\mu^2 < 0$ and $\lambda > 0$ are real-valued constants while $\bar{V} = 3 \,\mu^4 / {(2 \lambda)}$ \cite{Dehnen1}. The Higgs field in the spontaneously broken phase of symmetry leads to $v^2= \phi_0\,\phi_0^{\dagger}= -\, {6\mu^2}/{\lambda}$ and can further be resolved as $\phi_0=v N$ (where $N$ is a constant which satisfy $N^\dagger N=1$) with the introduction of the unitary gauge \cite{Dehnen2}. The general Higgs field $\phi$ may then be redefined through a real-valued excitation ($\xi$) of the Higgs scalar field as $ \phi=v \,(1+ \xi)^{1/2} \,N $.\\
The Higgs field in the broken phase of symmetry in this model possesses a finite range $L$ (i.e. Compton wavelength) which in the natural (geometric) system of units is inverse of the Higgs field mass of the model. Further, given the structure of this mass, it is unlike the usual Higgs mass of the SM,
\begin{equation}
M  =\left[ { \frac{1 + \frac {4 \pi }{3 \gamma } } {16 \pi
G ( \mu^4 / \lambda ) } } \right]^{-1/2}\,,
\end{equation}
with the gravitational constant $G=1/(\gamma v^2)$ where $\gamma \gg 1$ is defined as the square of the ratio of the Planck $(M_P)$ and gauge-boson $(M_A)$ masses \cite{Bezares07-BH,Dehnen3}. However, the effective gravitational coupling is given in terms of the scalar-field excitations, and it is defined as $\tilde {G} = {G}/({1+ \xi})$. In the absence of the scalar-field excitations, $G$ is automatically reproduced. With these considerations, the field equations corresponding to the action (\ref{action}) acquire the following form,
\begin{equation}
\nabla_\mu \partial^\mu \xi + M^2\xi \, = \, \frac{8\pi
\,G}{3} \,\hat{q}\, T,\label{Higgs1}
\end{equation}
\begin{equation}
{(1+\xi)} \, \left( R_{\mu\nu} - \frac {{R}}{2} \, g_{\mu\nu}
\right) + M^2 \xi \, \left( 1 + \frac {3}{4}\,  \xi \right)
g_{\mu \nu} \, + \nabla_\nu \, \partial_\mu \xi  = -8\pi G\, \left(
T_{\mu\nu} - \frac {\hat{q}}{3} T g_{\mu\nu} \right),\label{gravity}
\end{equation}
 \maketitle
where $T$ is the trace of the symmetric energy--momentum tensor $T_{\mu \nu}$ which belongs to the matter Lagrangian (${\cal{L}}_M) $ in the action given by equation (\ref{action}). Here $\hat{q}$ denotes the coupling of Higgs field to fermions. The conservation law $\nabla_\mu T^{\mu \nu}=0$ is satisfied for the case when $\phi$ does not couple to the fermionic state $\psi$ denoted as $\hat{q}=1$ in ${\cal{L}}_M$ \cite{Cervantes95a,Bezares07-DM}. In this case, the Higgs particles couple weakly and would rather be stable and thus improbably decay into less massive particles. For the the case when the Higgs field couples with the fermions (i.e. $\hat{q}=0$), the scalar field itself decouples in (\ref{Higgs1}), which means that the particles responsible for mass of elementary particles interact only gravitationally \cite{Cervantes95a,Bezares07-DM}.\\
The Ricci scalar may now be derived by calculating the trace of (\ref{gravity}). It reads as follows,
\begin{align}
  R=3M^2 \xi+ 8\pi \tilde{G} (1- \hat{q}) T.\label{Trace}
\end{align}
It is evident from equation (\ref{Trace}) that the present formulation in the limit $\xi \to 0$ and $T=0$ would consistently lead to all the physical consequences drawn by GR so far. However, for $\xi \neq 0$, there exists a cosmological function $\Lambda(\xi)=  8 \pi\, \tilde{G} \,V(\xi)$ (with $V(\xi) = [ 3 \,\mu^4 / {(2   \lambda)}]\, \xi^2$) which acts as a quintessential factor as discussed in forthcoming sections.

\section{Solutions in the limit of massless Higgs field}
\label{shiggs}

\begin{figure}[h] \centering
\includegraphics*[width=16.5cm]{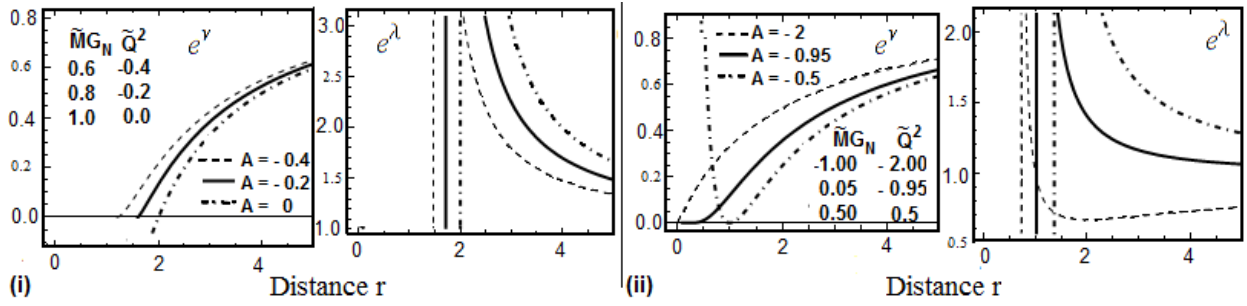}
\caption{Evolution of the metric components (up to second order in $r^{-n}$) for different negative values of $A$ with $r_S=B=2$ and $M_1G_N=1$. The effective Schwarzschild radius ($\tilde{M}$) diminishes (augments) for decaying (augmenting) values of $A$ (left side up). Quintessential attraction (cf. text) for $e^\lambda$ with higher values of $A$ is present ((ii) right).}
\label{fig1}
\end{figure}

\noindent In order to solve the field equations (\ref{Higgs1}) and (\ref{gravity}) explicitly and analytically, we will use a spacetime with spherical symmetry in the statical case, i.e. time derivatives of $\nu$, $\lambda$ and $\xi$ equal to zero. Let us consider, the following line element for our purpose,
\begin{align}
ds^2=e^{\nu (r)} \,dt^2- e^{\lambda(r)} dr^2- r^2 \,(d \vartheta^2 + sin^2 \vartheta \, d\varphi^2)\,.
\label{element}
\end{align}
Further, for simplicity, we assume phenomenologically that the matter represented by the Lagrangian in (\ref{action}) is a perfect fluid with the following energy--momentum tensor,
\begin{equation}
T_{\mu \nu} = \,(\epsilon + p)\, u_\mu u_\nu - p \, g_{\mu \nu}; \, \, \, \, \, u_\mu u^\mu =1 \, ,\label{a}
\end{equation}
where $u_\mu$ is the four-velocity of the fluid. A simple equation of state $p = w \,\epsilon$ can be considered. Such is generally followed by any perfect fluid relevant to the cosmology where the equation-of-state parameter $w$ (i.e. the ratio of the pressure $(p)$ to the energy density $\epsilon \equiv \varrho$ as $c=1$) is a constant which is independent of time.\\
In case of a point--mass in vacuum (or equivalently for the outer region of a massive object) and suppressing gauge fields, the Einstein equations (\ref{gravity}) in the component form now read as follows,
\begin{equation}
\frac{\nu''}{2}+\frac{\nu'^2}{4}- \frac{\nu'\lambda'}{4}-
\frac{\lambda'}{r}+ {M^2}\, \bar {\xi}\,  e^{\lambda} = F_1(\xi),\label{E1}
\end{equation}
\begin{equation}
\frac{\nu'}{2}- \frac{\lambda'}{2} + \frac{1}{r} (1-e^{\lambda})+ {M^2} \bar {\xi}\, r \,e^{\lambda} = F_2(\xi), \label{E2}
\end{equation}
\begin{equation}
\frac{\nu''}{2}+\frac{\nu'^2}{4}- \frac{\nu'\lambda'}{4} + \frac{\nu'}{r}+ {M^2}\, \bar {\xi}\,e^{\lambda} =  F_3(\xi), \label{E3}
\end{equation}
where $\bar {\xi}= (2 \xi + {3 \xi^2})/{4(1+\xi)}$. In equations (\ref{E1})--(\ref{E3}), prime denotes the derivative with respect to $r$ while the functions $F_1(\xi)$, $F_2(\xi)$, and $F_3(\xi)$ are defined as follows,
\begin{equation}
  F_1(\xi)=  \frac{1}{1+\xi}\left(
\frac{\lambda'}{2}\xi'-\xi''\right), \label{F1}
\end{equation}
\begin{equation}
  F_2(\xi)= - \frac{\xi'}{1+\xi}\,,\label{F2}
\end{equation}
\begin{equation}
  F_3(\xi)= -\frac{\xi'}{1+\xi}\,\frac{\nu'}{2}.\label{F3}
\end{equation}
These functions (\ref{F1})--(\ref{F3}) vanish when $\xi=const$. Further, the equation for the excited scalar field (i.e. a massive Klein--Gordon equation) given by (\ref{Higgs1}) in case of vacuum can be rewritten in form of a  Sturm--Liouville-type equation as below,
\begin{equation}
\left(\, r^2\, e^{(\nu-\lambda)/2} \, \xi'\, \right)' - r^2 \, M^{2}\,  e^{(\nu + \lambda)/2}\,  \xi = 0. \,  \label{Kge1}
\end{equation}
The gravitational-potential components in (\ref{Kge1}) emerge as a consequence of the covariant derivative which relates $\xi'$ to $\nu$ and $\lambda$. It is also worth noticing that for vanishing functions (\ref{F1})--(\ref{F3}) and Higgs field mass $M$, the equations (\ref{E1})--(\ref{E3}) would reduce to usual Einstein equations of GR and therefore automatically reflect the usual features of a Schwarzschild BH. Furthermore, a linear behaviour with $\hat{q}=1$ clearly shows that $M\to 0$ leads to the coupling $G$ related to the usual Newton's constant $(G_N)$ as $G=3G_N/4$ (cf. Section \ref{shiggs2}).\\
Now with the point--mass at $r=0$, the first integral of the excited Higgs field (i.e. the massless Klein--Gordon equation (\ref{Kge1}) with $M \to 0$ can be integrated once) is as follows,
\begin{equation}
\xi'=\frac{A}{r^2}\, e^{ (\lambda- \nu)/2}. \label{H1}
\end{equation}
The na\"ive properties of general Higgs fields depend ultimately on the coupling actually observed. The choice $M = 0$, however, cannot truly be expected in view of the SM of particle physics where the Higgs field definitely has a finite mass. Yet, in view of the cosmological properties of such family of fields within non-minimally coupled models, there are hints towards large length scales of the field (viz \cite{Bezares07-DM}) so that $M \to 0$ becomes at least a relevant analysis of behaviour as well as of basic properties. We will address the issue of a finite Higgs field mass with weak field limit in Section IV in detail.\\
The dimensionless integration constant $A$ in equation (\ref{H1}) is related to the properties of the gravitational object. As within GR, they act on outer fields by means of boundary conditions at the surface. Hence, $A$ is related to a mass $M_1$ of the massive body and more exactly on the integral over all density distributions of the gravitational source (which may differ from the latter). Here, we define the form of the integration constant $A$ in the asymptotic limit $r\rightarrow \infty$ as follows for the case $\hat{q}=1$ (see {\bf Appendix I}),
\begin{equation}
A= \,- \frac{2}{3} \, G\int T \sqrt{-g} \, d^3 x \, .\label{a}
\end{equation}
$A$ relates to a change in the strength of field excitations. Equation (\ref{a}) may be reduced to the following form which is valid for $\hat{q}=1$,
\begin{align}
A= \,\Omega \int \varrho\, \sqrt{-g} \, d^3 x \, ,\label{a1}
\end{align}
where $\Omega  \sim G\,(3w-1)$ in the case of a homogeneous mass sphere (which would be a good approximation for large distances $r$ in relation to the radius $R_1$ of the sphere with the mass given by the integral over energy-density distribution as $M_1=4 \pi R_1^3 \epsilon/3$). With $w = 0$, $A$ represents the situation as of pressure-less ordinary stars and galaxies without important inner structure (dust). $w = 1$ will represent  gravitational objects (usually highly relativistic) with an important structure represented by their (ideal) pressure term (stiff matter). $w=1/3$ shall not be a valid value for analysis since in that case there would be $A=0$ (i.e. $T=0$) and GR would be valid with $G=G_N$. However $w=-1$ represents anti-stiff matter  (i.e. quintessential structure). Within minimally coupled scalar-field theories with a quartic self-interaction, anti-stiff cosmologies with (effective) $w<-1$ are possible mainly due to quantum effects \cite{Onemi04,Onemi07}.\\
Subtraction of (\ref{E1}) from (\ref{E3}) leads to $\lambda'+ \nu'= F_3 (\xi) - F_1(\xi)$. Using equation (\ref{Kge1}) in the limit $M \to 0$, we obtain $F_1(\xi)+F_3(\xi)=  2 \xi'/ [r\,(1+\xi)]$ (see {\bf Appendix II}) which vanishes identically for vanishing excitations. However, equation (\ref{E3}) with some rearrangements in form of a total differentiation leads to
\begin{equation}
  \nu'= \frac{B}{r^2} \, \, \frac{1}{1+ \xi} \, \, e^{(\lambda- \nu)/2}\,, \label{nuprime}
\end{equation}
where $B$ is an integration constant which is related to the Schwarzschild mass. As in GR (where $p$ is also coupled to the gradient of the gravitational potential), the integration constant may possess pressure terms which may contribute to a dynamical mass other than the actual mass derived from density.\\
Let us further define a function $\tilde{g}(r)\equiv \tilde{g}$ as given  below,
\begin{equation}
\frac{1} {1+ \xi} \, \, e^{(\lambda- \nu)/2} = r \,\tilde g, \label{lambda-nug}
\end{equation}
such that an Abelian or a Riccati-type differential equation
is constructed for $\lambda'- \nu'$ in terms of $\tilde g$ in the following form,
\begin{equation}
\tilde g' = -\frac{1}{r}\left[\frac{AB}{2} \tilde g^3+ (2A + B) \, \tilde g^2+ \tilde g\right]. \label{g'K}
\end{equation}
With these considerations, the excited Higgs field given by equation
(\ref{H1}) leads to $\xi = e^{A\nu/B} - 1$. A transparent analytical approach for such fields is  quite difficult in view of the exact solution of (\ref{g'K}), and further purely numerical considerations may not help in the task of understanding and interpreting the impact of these effects  on usual astrophysical or solar-relativistic calculations. We have therefore looked for some approximated solutions by using a series-expansion method. Such analysis shows the basic properties of the scalar-field interaction (and self-interactions) for small but non-vanishing excitations of the scalar field, which is especially valid for large distances to the source of gravitation as well as for the bodies having relatively low mass. Let us consider the following series as an ansatz for $\tilde g$ to further simplify the equations (\ref{E1})--(\ref{Kge1}),
\begin{equation}
\tilde g = \sum_{n=1}^\infty \frac{C_n}{r^n} =  \frac{1}{r}\left[C_1 + \frac{C_2}{r}+ \frac{C_3}{r^2}+
  \frac{C_4}{r^3}+\, \cdots \,\right].\label{gtilda}
\end{equation}
Using equation (\ref{gtilda}) in (\ref{g'K}) and then comparing
the left- and right-hand sides of equation (\ref{g'K}) itself, the coefficients of $r^{-1}$, $r^{-2}$ and so on can be obtained with the following simple recursion relations (which are reported here only up to the fifth order of $1/r$) with straightforward calculations as follows,
\begin{equation}
\left.
\begin{array}{c}
C_1 =1\\
C_2 =2A + B\\
C_3 = (2A + B)^2 + \frac{AB}{4}\\
C_4= (2A + B)^3 + \frac{2 A B}{3} \, (2A + B)\\
C_5= (2A + B)^4 + \frac{29 AB}{24}\, (2A + B)^2 + \frac{3 (AB)^2}{32}
\end{array}
\right\}.\label{recur}
\end{equation}
Clearly, the constants $C_i$ appear as additive and multiplicative terms of $A$ and $B$, and these are the only two parameters of physical interest of the present model. Consequently, we restructure (\ref{gtilda}) as follows,
\begin{equation}
  \tilde{g}= \frac{1}{r}\left[1- \frac{(2A + B)}{r}\right]^{-1}+\, \frac{A
    B}{2 r}\,\, X(A,B;r^{-n});~~~~~~~~~~~~~~~ (n\geq 2)\,,\label{gtilda2}
\end{equation}
where $X(A,B;r^{-n})$ is a function of $r$, $A$ and $B$ only, as given below,
\begin{equation}
  X(A,B;r^{-n})= \frac{1}{2 r^2} \, + \, \frac{4\,(2 A+ B)}{3 r^3} \, + \, \frac{1}{4 r^4} \, \left[
  \frac{58\, (2A +    B)^2}{3} \, + \, \frac{3\, AB}{2} \right] \, + \,\cdots\,;~~~~~~ (n\geq 2).\label{X-term}
\end{equation}
In fact, with $n\geq 2$, $X(A,B;r^{-n})$ is negligible for extremely large distances. Low potency terms appear as small corrections for smaller distances to the gravitational ``source''. Such a situation can physically be understood in terms of the weakening of gravity once one moves far away from the centre of a gravitating mass.  Now subtracting the equation (\ref{E1}) from equation (\ref{E3}) and using the equation (\ref{nuprime}), we have
\begin{equation}
  e^{(\lambda+ \nu)/2}= (1+ \xi)\left[\frac{1}{r \tilde{g}}+ \frac{2A + B}{r}+ \frac{AB}{2r} \tilde{g}\right].\label{lambda+nu}
\end{equation}
Using the equations (\ref{gtilda2}) and (\ref{lambda+nu}) further leads to
\begin{alignat}{1}
    e^\lambda=& \left[1 - \left(\frac{2 A + B}{r}\right)\right]^{-1} - \frac{AB}{2 r^2}\left[1- \frac{(2 A + B)}{r}\right]^{-2}+\label{lambda1}\\
     &+\frac{AB}{2 r} (2 A+ B) X(A,B;r^{-n})- \left(\frac{AB}{2 r^2}\right)^2 Y(A,B;r^{-n})\,,\nonumber
\end{alignat}
where $Y(A,B;r^{-n})$ is a function of $X(A,B;r^{-n})$ itself. The equation (\ref{lambda1}) may further be rewritten into the following form,
\begin{alignat}{1}
  e^\lambda=& \left[1- \frac{(2 A + B)}{r} - \frac{AB}{2 r^2}\right] \left[1- \frac{(2 A + B)}{r}\right]^{-2}+ \label{elambdac}\\
   &+\frac{AB}{2 r} (2 A+ B) X(A,B;r^{-n})- \left(\frac{AB}{2 r^2}\right)^2 Y (A,B;r^{-1})\,,\nonumber
\end{alignat}
which after some straightforward simplifications leads to the following form of the equation (\ref{elambdac}),
\begin{equation}
  e^\lambda= \left[1- \frac{(2 A+ B)}{r}+ \frac{AB}{2 r^2}\right]^{-1}+ \frac{AB}{2 r}(2 A+ B)X(A,B;r^{-n})- \frac{(AB)^2}{4 r^2} Z(A,B; r^{-n})\,,\label{lambda2}
\end{equation}
where $Z(A,B; r^{-n})$ is a function of $Y(A,B; r^{-n})$ or ultimately a function of $X(A,B; r^{-n})$. For the potency $n$, there is again $n\geq 2$. Finally, up to the second order in $1/r$, the equation (\ref{lambda2}) yields
\begin{equation}
  e^\lambda= \left[1- \frac{2\tilde{M}G_N}{r}+ \frac{\tilde{Q}^2}{r^2}\right]^{-1}\,,\label{elambda}
\end{equation}
with the effective mass parameter and a RN-like charge term as defined in the following form,
\begin{equation}
  \tilde{M} G_N= A + \frac{B}{2}, \label{mass}
\end{equation}
\begin{equation}
  \tilde{Q}^2= \frac{AB}{2},\label{charge}
\end{equation}
where $G_N$ is the Newtonian gravitational coupling constant. The effective mass, as a general dynamical mass and in contraposition to the ``actual'' mass, is dependent on the one which is derived from energy density $(\epsilon)$ itself as well as on pressure $p$ which enters a measured mass term through the integral of the trace of the energy--momentum tensor $T$. Further, the generalised charge parameter $\tilde{Q}$ appearing as a consequence of the usual gravitational terms hidden in $B$ coupled to the gravitational scalar-field terms in $A$, may act against the usual gravitation of GR in the same way as the RN charge does for a charged point--particle in a gravitational field \cite{Rei16}. This quintessential behaviour grows for higher values of the ``charges'' (i.e. higher masses or strong field excitations) and smaller distances to the gravitating body. The physical properties of the solutions are visible, although the exact vanishing of horizons \cite{Bezares07-BH} is due to at-low-distance dominant terms which appear for high-order corrections. It is also possible to interpret these terms acting in the dynamics by using the weak-field behaviour, for instance for the  solar-relativistic calculations which are dealt separately \cite{Bezares09-DM}. They need of such second-order components of the metric result,
\begin{equation}
  e^\nu= \left[\frac{\left(1- \frac{2\tilde{M}G_N}{r}\right)^2\, e^\lambda}{\left[1- \left(1- \frac{2\tilde{M}G_N}{r}\right) \tilde{Q}^2X(A,B;r^{-n})\right]^2}\right]^\frac{B}{2\tilde{M}G_N}\,;~~~~~~~~~~ (n\geq 2).\label{enunu}
\end{equation}
Using (\ref{lambda2}) and with some straightforward calculations, the equation (\ref{enunu}) leads to the following form,
\begin{equation}
  e^\nu= \left\{\frac{1- \frac{2\tilde{M}G_N}{r}- \frac{\tilde{Q}^2}{r^2}+ \left(1- \frac{2\tilde{M}G_N}{r}\right)^2 \left[\frac{\tilde{Q}^2}{r}2\tilde{M}G_N X(A,B; r^{-n})- \frac{\tilde{Q}^4}{r^2}Y(A,B; r^{-n})\right]}{\left[1- \left(1- \frac{2\tilde{M}G_N}{r}\right)\tilde{Q}^2 X(A,B; r^{-n})\right]^2}\right\}^{\frac{B}{2\tilde{M}G_N}}\,,
\end{equation}
which may then further be restructured for $n\geq 2$ as follows,
\begin{alignat}{1}
  e^\nu=& \left\{\left[1- \frac{2\tilde{M}G_N}{r}- \frac{\tilde{Q}^2}{r^2}+ \frac{\tilde{Q}^2}{r}\left(1- \frac{2\tilde{M}G_N}{r}\right)^2\left(2\tilde{M}G_N X(A,B; r^{-n})- \right.\right.\right.\nonumber\\
  &\left.\left.\left.- \frac{\tilde{Q}^2}{r} Y(A,B;r^{-n})\right)\right]\left[1+ \frac{\left(1- \frac{2\tilde{M}G_N}{r}\right)\tilde{Q}^4 X(A,B; r^{-n})^2}{1- \left(1- \frac{2\tilde{M} G_N}{r}\right) \tilde{Q}^2 X(A,B; r^{-n})}\right]^2\right\}^{\frac{B}{2\tilde{M} G_N}}\,,\label{enuxx}
\end{alignat}
Unlike in the equation (\ref{elambda}) for $e^{\lambda}$, up to second order the generalised charge parameter $\tilde{Q}^2$ cancels out in (\ref{enuxx}) and the metric component $e^{\nu}$ thus evolves as given below,
\begin{equation}
  e^\nu= \left[1- \frac{2\tilde{M}G_N}{r}\right]^\frac{B}{2\tilde{M}G_N}\,.\label{enu2c}
\end{equation}
In linear approximation, from (\ref{enu2c}) we have $e^\nu= 1- B/r$ with $B$ as a dynamical mass parameter (see linear approach). For $A\neq 0$, the power coefficient $B/(2\tilde{M}G_N)$ may be written as $(1+ 2A/B)^{-1}$, thus showing the deviation from a usual Schwarzschild value of $e^\nu$ (with $B=r_S$ as Schwarzschild radius) as discussed later. Even for weak gravitational fields, a non--vanishing scalar field appears related to the density and pressure terms as a dynamical correction to the bare mass $M_1$ (see next Section for details). For $A/B\ll 1$, for relatively weak field regimes, there is clearly a RN-like solution for $\lambda$ as given in (\ref{elambda}) with a generalised charge parameter $\tilde{Q}$. For $\nu$ in (\ref{enu2c}), a quadratic term in $r$ may only appear as consequence of the potency term (i.e. from the relation between the amplitude of (\ref{nuprime}) and the effective mass parameter (\ref{mass})). An effective mass appears from an analogy to the Schwarzschild solution which depends on scalar-field contributions related to the pressure $p$.\\
The excitation of the scalar field for small mass $\tilde{M}$ and charge $\tilde{Q}$ in relation to the distance (i.e. beyond high-field regimes) now reads as follows,
\begin{equation}
  \xi = \left[1- \frac{2\tilde{M}G_N}{r}\right]^\frac{A}{2\tilde{M}G_N} - 1\,.\label{xiq}
\end{equation}
It is exactly vanishing for the Schwarzschild metric ($A=0$). However, for a RN-like solution ($A\neq 0$), there is $\xi\gtrsim 0$ for the typical value $\tilde{M}>0$ with $A<0$. Thus for a vanishing excitation parameter $A$, the Schwarzschild metric is valid. Negative values of $A$, on the other hand, lead to a positive field with a singular value at $r=0$ and the tendency $\xi\rightarrow 0$ for spatial infinity.\\
The metric component  $e^\lambda$ (\ref{elambda}) shows a RN-like form and up to second order vanishes for
\begin{equation}
  r_\pm = \tilde{M} G_N \pm \sqrt{(\tilde{M}G_N)^2- \tilde{Q}^2}\,.\label{rpm}
\end{equation}
Given the vanishing of horizon for $A<0$ \cite{Bezares07-BH}, it has a regime where the validity of approximation clearly breaks down. However, a changed behaviour from usual quasi-Schwarzschild character for an almost flat metric towards the vanishing of the singularity for the exact solution is present. Second-order approximation has indeed a Reissner--Nordstr\"{o}m character and thus pretends following three cases: {(i)} - extremal BH when $(\tilde{M}G_N)^2=\tilde{Q}^2$ (for which the concentric event horizon becomes degenerate), {(ii)} - a naked singularity with $(\tilde{M}G_N)^2<\tilde{Q}^2$, and {(iii)} - a Schwarzschild case for $(\tilde{M}G_N)^2>\tilde{Q}^2$.  The case (iii) also appears when the field excitations $\xi$ vanish completely (for which $\tilde{Q}$ is zero exactly when $A=0$), which is clear from the equations (\ref{elambda}), (\ref{enu2c}) and (\ref{xiq}) respectively.\\
\noindent Within RN approximation for $A<0$ and $B>0$, only the case (iii) is indeed possible ($\tilde{Q}^2<0$), leading to a quasi-Schwarzschild behaviour for low-field regimes. Nevertheless, the analogy to RN solutions is an interesting subject which reminds that for a massive object whose charge ($AB$) is not neutralised by further effects, the Schwarzschild radius itself loses its meaning of dominant property of the system. Here, the radius given by the generalised charge ($r_Q\equiv \sqrt{|\tilde{Q}^2|}$) is an intrinsic quality which affects the Schwarzschild radius itself, and the weakening of the gravitational fields in weak field limit appears indeed as a consequence of the correction terms. In Fig. \ref{fig1}, this can be seen especially for relatively high-order terms of $-A$. Further, $e^\lambda$ becomes flatter as the effective mass tends to zero. For low gravitational regimes, an effective, dynamical mass replaces the role of bare, luminous mass, and the rise of quintessential terms is thus clear. 
\noindent Considering this fact into account, it might be possible to establish measurably relevant distinctions of this induced-gravity model to usual dynamics even at long-scale regimes such as those of galactic bulges as well as relevant indications for intermediate regimes towards strong gravitational fields.\\
\noindent  The case $A\leq -1$ is especially interesting. Dynamically speaking, such systems with $M_1>0$ and $\tilde{M}<0$ are \emph{Quintessentially Attracted}. In fact, these systems are quintessential because of the anti-gravitational behaviour of $e^\lambda$ following the negative effective (yet positive actual) mass. They are attractive because of the behaviour of $e^\nu$ which still shows attraction of the gravitating body lying at $r=0$. However, the Schwarzschild radius is vanishing. In this analysis, the role of the ``charge'' radius $r_Q$ is important. Detailed thorough discussion, though, needs of the values of $A$ and $B$ in terms of mass and pressure (cf. next Section). This is only possible in relation to a linear analysis for interpretation (in the same way that the coupling constant $\kappa$ of GR is related to $G_N$ by means of comparison with the weak-field behaviour). The higher-order corrections are also relevant for considerations near as well as beyond the Schwarzschild and charge radius. The evolution of the metric components for the negative values of $A$ as presented in Fig. \ref{fig1} shows that the effective Schwarzschild radius augments for higher values of $A$. The positive values for $A$ also lead to similar results. \\
\noindent Further, if we consider positive amplitudes $A>0$ (with $B>0$), the gravitational field is strengthened and the effective Schwarzschild radius shifts to $r>2M_1G_N$ such that the gravitational attraction becomes greater as related to a relatively higher dynamical mass $\tilde{M}>M_1$. We have observed that in this case the scalar-field excitation leads to a strengthening of the gravitational coupling and may thus be of special relevance in terms of Dark Matter. The presence of a scalar field for Quintessence generally changes the singularity of Black Hole solutions \cite{Wetterich01}, and further, models of Quintessence usually predict long-range forces mediated by the fields leading to different concepts of effective mass \cite{Wetterich03}, especially as fluctuations of the scalar field which may behave similarly to relativistic gas \cite{Wetterich02} and/or be associated to halo mass of galaxies \cite{Wetterich01}.\\

\section{Solutions with Massive Higgs field in weak field limit}
\label{shiggs2}
\noindent For a comprehensive description  of the appearing factors, fields and relations, we analyse the linear field equations of the present model. Such analysis is not only important for the correct  physical interpretation of the gravitational potential and the effective mass, but also useful to analyse the galactic dynamics at distances as galactic bulges and spirals \cite{Bezares07-DM, Cervantes07a}. \\
The mass of the scalar field would gain relevance at distances near to the length scale $L$.  It seems then obvious to use the linear equations with finite values of $M$ and to a-posteriori define $M$ to be high-valued in accordance to the previous section. With the approximation near to the Minkowski flat metric such that $\lambda \ll 1$, $\nu \ll 1$ and  $\xi \ll 1$ (i.e. the static linear approximation), the in-vacuo statical Einstein's equations  in component form read as given below,
\begin{equation}
\frac{\nu''}{2} - \frac{\lambda'}{r} + \xi'' = - \frac{M^2}{2} \xi  ,\label{E11}
\end{equation}
\begin{equation}
\frac{\nu'}{2}- \frac{\lambda'}{2} - \frac{\lambda}{r} + \xi' = - \frac{M^2} {2} {\xi}\, r, \label{E22}
\end{equation}
\begin{equation}
\frac{\nu''}{2}+ \frac{\nu'}{r} = -\frac{M^2}{2} {\xi}.  \label{E33}
\end{equation}
Neglecting an anti-Yukawa solution, the general linear vacuum solutions of $\xi$, $\nu$ and $\lambda$ read
\begin{equation}
  \xi=\frac{C}{r}e^{-M r},\label{xe}
  \end{equation}
 \begin{equation}
  \nu= -\frac{C}{r}e^{-M r}- \frac{D}{r}, \label{de}
\end{equation}
\begin{equation}
  \lambda= -\frac{C}{r} \left(1+ 2 M r\right) e^{-M r} + \frac{D}{r}+ F\,, \label{dee}
\end{equation}
where  $C$, $D$ and $F$ are the integration constants. In (\ref{dee}), $F=0$ given the condition $\lambda\rightarrow 0$ at spatial infinity. Further, it may be easily seen that $C$ is related to the constant $A$ of the exact solution with $M\to 0$ while the constants $C$ and $D$ are related to $B$ as discussed in the previous section. The form of $e^\nu$ and $e^\lambda$ in (\ref{de}) and (\ref{dee}) is related to the asymptotic value of $B/r$ for $e^\nu$ and $(2A+B)/r$ for $e^\lambda$ respectively as presented by $\tilde{g}$ in the last section.\\
The exact physical meaning of these integration constants can be derived by solving the inhomogeneous equations, i.e. the linearised equations (\ref{E11})--(\ref{E33}) in the presence of the source. Examples are -- a polytropic density distribution \cite{Bezares07-DM}, a Freeman-disk profile \cite{Cervantes3} or a homogeneous mass distribution which then gives the solution for a point--particle when the radius $R_1$ of the gravitating body is taken as $R_1\rightarrow 0$.\\
For an ideal fluid case with (\ref{a}), and for the inner space (i.e. $r<R_1$) \cite{Bezares07-DM}, the sources of equations (\ref{E11})--(\ref{E33}) read respectively as follows,
\begin{align}
  -\frac{8\pi G}{3}\left(2\epsilon+ 3p\right),\,\,\,-\frac{8\pi G}{3} \epsilon\, r;\,\,\,\frac{8\pi G}{3} \epsilon.
\end{align}
The scalar-field equation (\ref{Kge1}), however, reads as below,
\begin{align}
  \xi''+ \frac{2}{r}\,\xi'- M^2\xi= -\frac{8\pi G}{3}(\epsilon- 3p).\label{sfes}
\end{align}
As within GR, the integration constants (in equations (\ref{xe})--(\ref{dee})) are related to effective mass terms entailing density terms of the inner matter. Mass is related to density by means of its integration over volume. For solving the system of integration constants in (\ref{xe}) and (\ref{dee}), we may solve the scalar-field equation for $r<R_1$ to find $C$ by means of boundary conditions at the surface $R_1$ of the gravitational object and then the one for $\nu$ to obtain $D$. Further, we have the following Poisson equation which is valid for a potential $\Psi= \Phi+ \xi/2$ and $\Phi=\nu/2$ (see \textbf{Appendix III}),
\begin{equation}
  \nabla^2 \Psi= 4\pi G(\epsilon+ 3p). \label{Poisson}
\end{equation}
It shows the Newtonian character of the potential $\Psi$ and the gravitational properties of $\xi$ as a Yukawa potential.\\
We will assume large distances $r$ in relation to the radius $R_1$ of the gravitational object so that the solution for a point--particle with inner structure (i.e. considering the pressure which is related to the scalar-field excitation amplitude) will be given. The Dark Matter profile for exactly flat rotation curves within this model was recently analysed \cite{Bezares09-DM}, and consequences of such pressure terms can be investigated by using the corresponding Poisson equation.\\
The inner solution of the scalar-field excitation leads in the case of $R_1\ll r$ and without a point--masses at the centre to the integration constant $C$ as follows for a point--particle (see \textbf{Appendix III}),
\begin{equation}
  C= \frac{2M_1 G_N}{4}(1- 3w),\quad (R_1\ll r)\,.\label{Cconst}
\end{equation}
For $R_1M\ll 1$, we obtain (see \textbf{Appendix III}),
\begin{align}
  D=\frac{1}{2}M_1G_N(1+ 3w),
\end{align}
According to equation (\ref{Poisson}), $\Psi= -(M_1G/r)(1+ 3w)$ gives the Newtonian potential in the absence of fields $\xi$. However, $\Phi$ gives the actual gravitational potential in the following form for point--particles and $r\gg R_1$ (see \textbf{Appendix III}),
\begin{equation}
  \Phi= -\frac{M_1 G_N}{r}\left(1+ \frac{3}{2}w\right),\label{Phib}
\end{equation}
which is valid for $M r \ll 1$, with a gravitational coupling $G=3G_N/4$. Clearly, $w$ adds to a dynamical mass as given below,
\begin{equation}
  M_{dyn}= \left(1+ \frac{3}{2}w\right) M_1; ~~~~~~~~~~~~~~~~(M R_1\ll 1),\label{mdyn}
\end{equation}
which is valid for $M R_1\ll 1$. In the general case, there is a usually small correction coming from $MR_1$-dependency of dynamical mass (see also \cite{Bezares09-DM}). Now, with such considerations, we have
\begin{equation}
  \nu=- \frac{2M_{dyn} G_N}{r}.\label{nudyn}
\end{equation}
This form is consistent with a parameterised Post-Newtonian (PPN) framework. Now from equation (\ref{enu2c}), it is clear that in linear approximation a metric component has the following form,
\begin{equation}
  e^\nu= 1- \frac{B}{r},\label{enub}
\end{equation}
where $B$ in its asymptotic limit can be written directly in the form given below by using the equation (\ref{Phib}),
\begin{equation}
  B= 2M_1 G_N \left(1+ \frac{3}{2}w\right).\,\label{Bconst}
\end{equation}
It defines a dynamical Schwarzschild radius $r_{dyn}$ which is to be compared to the effective radius $\tilde{r}_S$ in (\ref{mass}).\\
The linear approach is consistent with the series-expansion method as used above, and the potency term of (\ref{enu2c}) may be written as an effective-mass ratio as follows,
\begin{equation}
  \frac{r_{dyn}}{\tilde{r}_S}= \frac{B}{2\tilde{M}G_N}.
\end{equation}
In linear approximation for $M\rightarrow 0$, according to equation (\ref{a1}), $A$ is given as
\begin{equation}
  A=- \frac{1}{2}M_1 G_N(1- 3w).\label{Aconst}
\end{equation}
Hence, for weak-field regimes, $A$ is then equal to $-C$ as long as $M \to 0$. Further, for point--particles with $M r\ll 1$, linear approach leads to
\begin{equation}
  \lambda= h(w) \frac{2M_{dyn} G_N}{r},
\end{equation}
with a (lowly $Mr$-dependent) parameter $h(w)$ well-given by
\begin{equation}
  h(w)= \frac{1+ 8w}{2+ 3w}
\end{equation}
for $M\rightarrow 0$ and for non-dominant $\tilde{Q}$-charges. Here $h(w)$ is a term which shows for low pressures (and low distances compared to $L$) the ratio of effective masses in (\ref{mass}) and (\ref{Bconst}), the significance of which is especially visible for $w=0$. However, terms of $h(w)$ actually dependent on distance--length-scale relation (from a linear analysis for general scalar-field masses which we do not report here explicitly) are negligible for $rM\ll 1$, and derivatives of $h(w,r)$ are of the magnitude $M^2$.\\
As evident from equations (\ref{elambda}), (\ref{enu2c}), (\ref{Bconst}) and (\ref{Aconst}), for $w=0$, the linear approximation shows $\nu=-2\lambda$. A finite value of the parameter $w$ related to $A$ seems necessary within a PPN framework (see discussion on orbit dependence upon $w$ in Section V), and indeed, for $w=1/6$, there is exactly $\tilde{r}_S= 2M_1G_N$. For $w=1/5$, further, there is also $h(w)r_{dyn}=2M_1G_N$. Moreover, gravitational energy-density analyses (viz \cite{Bezares09-DM}) constraint $w$ to about one fifth to one sixth, and indeed, for $w=1/5$ there is $h(w)=1$, and the metric components valid in linear approximation have the following form,
\begin{equation}
  e^\nu= 1- \frac{2M_{dyn} G_N}{r},
\end{equation}
\begin{equation}
  e^\lambda= 1+ \frac{2M_{dyn} G_N}{r} .\label{lambdadyn}
\end{equation}
Furthermore, solar-relativistic effects can then be expected to be given as they are measured for all low-energy systems and with advances of perihelion dependent on the system's internal structure ($p$). For low gravitating systems, effective masses $\tilde{M}$ and $M_{dyn}$ are approximately the same and the dynamical mass $M_{dyn}$ takes the place of the actual mass $M_1$.\\
\begin{figure}[h] \centering
\includegraphics*[width=16.5cm]{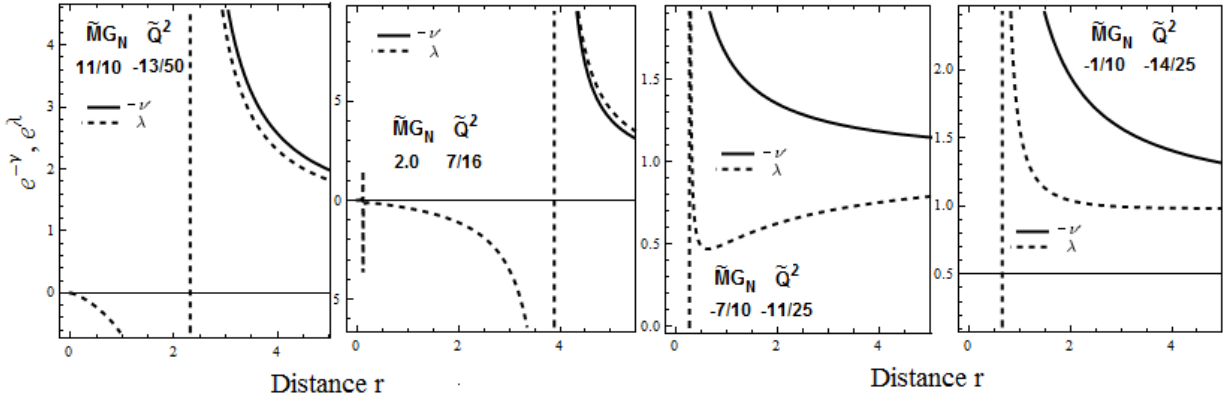}
\caption{Evolution of $e^{-\nu}$  and $e^\lambda$ for (i) $w=1/5$, (ii) $w=1/2$, (iii) $w=-2/5$ and (iv) $w=-1/5$ with $M_1G_N=1$. Stiff matter $w>1/3$ corresponds to $\tilde{Q}^2>0$. The dynamical mass for linear approximation reads $M_{dyn}= 13/10$ for $w=1/5$.}
\label{fig4}
 \end{figure}
\noindent Using (\ref{Bconst}) and (\ref{Aconst}), a closer look at
$e^\lambda$ and $e^\nu$ in dependence of $w$ may be taken into account. In second approximation, for positive pressures $p$ (Fig.\ref{fig4}, (i) and (ii)), the effective Schwarzschild radius decreases in respect to the one given by $M_1$, corresponding to the case $A<0$ as discussed earlier. For stiff matter $w>1/3$ ($A>0$), $e^{-\nu}$ has lower values than $e^\lambda$. For negative pressures, on the other hand (see Fig.\ref{fig4} (iii) and (iv)), $w<-1/6$ leads to quintessential attraction for $e^\lambda$. For $w<-2/3$ there is  also a gravitational repulsion. $\tilde{Q}^2$ is always smaller than $\tilde{M}$ unless for $w\lesssim -0.7$, for which $e^\nu$ is nearly flat.\\

\section{Effect of field excitations on geodesic motion}
\label{gm}
\noindent We now try to analyse the singularities discussed above in view of the completeness of geodesics corresponding to the line element (\ref{element}) with the metric components as mentioned in Section \ref{shiggs}. The geodesic motion especially in view of the RN-like charge parameter  arising because of the non-vanishing field excitations is investigated. We use $g_{\mu \nu}g^{\mu \nu}=-\varepsilon$ with $\varepsilon=0$ and $\varepsilon=-1$ to represent the constraints for
the null and timelike geodesics respectively. The geodesic equations corresponding to the metric (\ref{element}) are given as follows,
\begin{equation}
\ddot t = - \dot t\,\,  \dot r \, \, \nu', \label{g1}
\end{equation}
\begin{equation}
\ddot r = - \frac{1}{2} e^{-\lambda} \left ( -2 r \, (\dot \vartheta^2 +
\sin^2 \vartheta \, \dot \varphi^2) \, +  e^{\lambda}\,  \dot r^2 \, \lambda' +
 e^\nu \, \dot t^2 \, \nu' \right ), \label{g2}
\end{equation}
\begin{equation}
\ddot \vartheta  =  -\frac{2}{r}\,  {\dot r \dot \vartheta} + \cos \vartheta
 \sin\vartheta \, \, \, \dot \varphi^2, \label{g3}
\end{equation}
\begin{equation}
\ddot \varphi =  - \frac{2} {r}\, \dot r \dot \varphi  - \cot\vartheta \, \,
\dot \vartheta \, \dot \varphi  , \label{g4}
\end{equation}
where the dots and primes represent the differentiations with respect to
the affine parameter $\tau$ and $r$ respectively. Equation (\ref{g1}) has the solution $\dot t = E\, e^{-\nu}$, and using $\vartheta = \pi/2$ (equatorial plane) it leads to $\dot \varphi = L/r^2$ where $E$ and $L$ are integration constants. Now, using the constraint for timelike and null geodesics, we obtain
\begin{equation}
    v^2 = \left (\frac{d r}{dt} \right )^2 =  e^{-(\nu +\lambda)} E^2 \left (  1 - \frac{L^2
    \, e^{\nu} }{E^2 \, r^2} + \frac{\varepsilon \, e^{\nu} }{E^2} \right ). \label{g5}
\end{equation}
However, from the geodesics equations, we obtain the expression for tangential velocity as follows,
\begin{equation}
    \Omega = \frac{d \varphi}{d t}= \frac{1}{r^2}e^\nu \label{g51}
    \left(\frac{L}{E}\right).
\end{equation}
Using equations (\ref{g5}) and (\ref{g51}), one can write the angular velocity in the following form to describe the radial orbit changes,
\begin{equation}
    \left(\frac{d \varphi}{dr}\right) =  \frac{L}{E r^2} \,
    e^{\frac{(\nu + \lambda)}{2}} \left ( 1 - \frac{L^2 \, e^{\nu}}{E^2 \, r^2} + \frac{\varepsilon \, e^{\nu} }{E^2}  \right )^{-\frac{1}{2}}, \label{g6}
\end{equation}
\noindent Using (\ref{g5}) and first considering the RN-like
charge parameter such that $|\tilde{Q}^2|\ll r^2$, an effective
potential may be defined as following,
\begin{equation}
  V_{eff}= \frac{\varepsilon \tilde{M}G_N}{r}+ \frac{L^2}{2 r^2}- \frac{\tilde{M}G_N
  L^2}{r^3}\,. \label{vefff}
\end{equation}
The equations (\ref{g5}) and (\ref{vefff}) satisfy the following
energy law,
\begin{equation}
  {\cal E} =  \left(\frac{dr}{dt}\right)^2 + V_{eff}= \frac{1}{2}\left\{\frac{\left(1- \frac{2\tilde{M}G_N}{r}\right)}
  {\left(1-\frac{2\tilde{M}G_N}{r} \right)^{\frac{B}{2\tilde{M}G_N}}}\,E^2+ \varepsilon\right\}.\,\label{rd2}
\end{equation}
In equation (\ref{rd2}), for $A\ll B$, we have
\begin{equation}
  \frac{\left(1- \frac{2\tilde{M}G_N}{r}\right)}{\left(1-\frac{2\tilde{M}G_N}{r} \right)^{\frac{B}{2\tilde{M}G_N}}}= \left(1- \frac{2\tilde{M}G_N}{r}\right)^{\frac{2A}{B}- \frac{4A^2}{B^2}}\,.
\end{equation}
The effective potential (\ref{vefff}) has Newtonian form for
$r\rightarrow \infty$, and it possesses an extremal value for
\begin{equation}
  r= -\frac{L^2}{2 \varepsilon \tilde{M}G_N} \left[1\mp \sqrt{1+ \frac{3(2\tilde{M}G_N)^2 \varepsilon}{L^2}}\right]\,.
\end{equation}
The innermost stable circular orbit (ISCO) is then given by
$r=6\tilde{M}G_N$, which is related to $L/(\tilde{M}G_N)=\sqrt{12}$. For timelike geodesics, the maximal momentum ($L_x$)--mass relation for the extremum is then given by
\begin{equation}
  \frac{L_x^2}{(M_1G_N)^2}= 3 (1+ 6 w)\,.
\end{equation}
\begin{figure}[h] \centering
\includegraphics*[width=16.5cm]{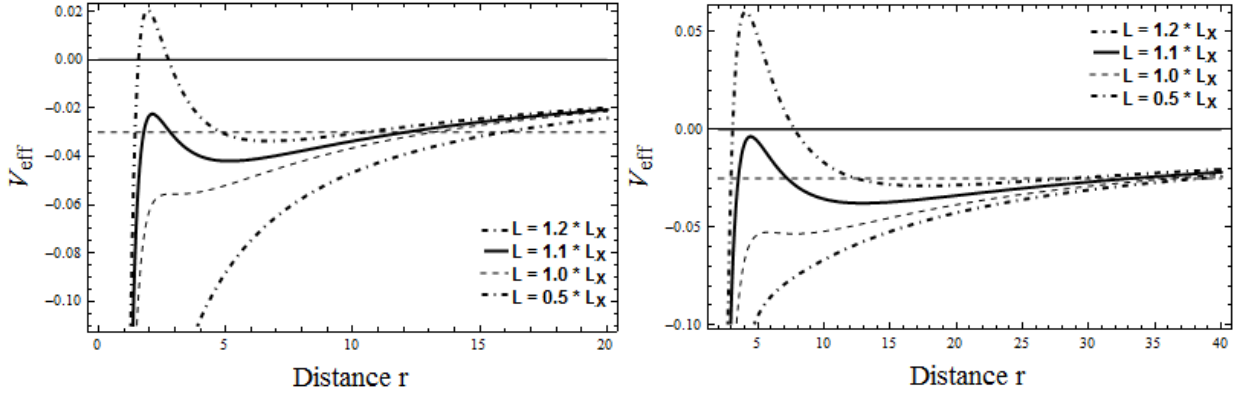}
\caption{Effective potential with (i) $w=0$ (left panel), and (ii) $w=0.2$ (right panel) for different values of $L$ with $M_1G_N=1$. Bound states appear for high enough stiffness and momentum $L$ which oscillate between the perihelion and the aphelion  for ${\cal E}<0$. An orbit corresponding to maximum (minimum) energy ${\cal E}$ is unstable (stable).}
\label{fig6}
 \end{figure}

\noindent Fig.\ref{fig6} represents the dependence of effective potential on $w$, and such dependence is related to the difference between luminous and dynamical mass, tangential and angular velocity. It is therefore useful to analyse these mass and velocity parameters for different pressure terms. The tangential and angular velocities are obtained by using equations (\ref{g5}) and (\ref{g6}) along with (\ref{elambda}) and  (\ref{enu2c}) are given below,
\begin{equation}
  \left(\frac{dr}{dt}\right)^2= \left[1- \frac{2\tilde{M}G_N}{r}\right]^\frac{B}
  {2\tilde{M}G_N}\left[1- \frac{2\tilde{M}G_N}{r}+ \frac{\tilde{Q}^2}{r^2}\right]
  \left[1- \frac{ \left(1- \frac{2\tilde{M}G_N}{r}\right)^\frac{B}{2\tilde{M}G_N}
  \left(L^2- r^2\varepsilon\right)}{E r^2}\right]\,,
\end{equation}
\begin{equation}
  \left(\frac{d \varphi}{dr}\right)^2= \frac{L^2 \left[\left(1- \frac{2\tilde{M}G_N}{r}\right)^{\frac{B}{2\tilde{M}G_N}}+ \left(1- \frac{2\tilde{M}G_N}{r}+ \frac{\tilde{Q}^2}{r^2}\right)^{-1}\right]}{E^2 r^4 \left[1- \frac{\left(1- \frac{2\tilde{M}G_N}{r}\right)^{\frac{B}{2\tilde{M}G_N}}}{E^2 r^2} \left(L^2- r^2 \varepsilon\right)\right]}\,.
\end{equation}
\begin{figure}[h] \centering
\includegraphics*[width=16.5cm]{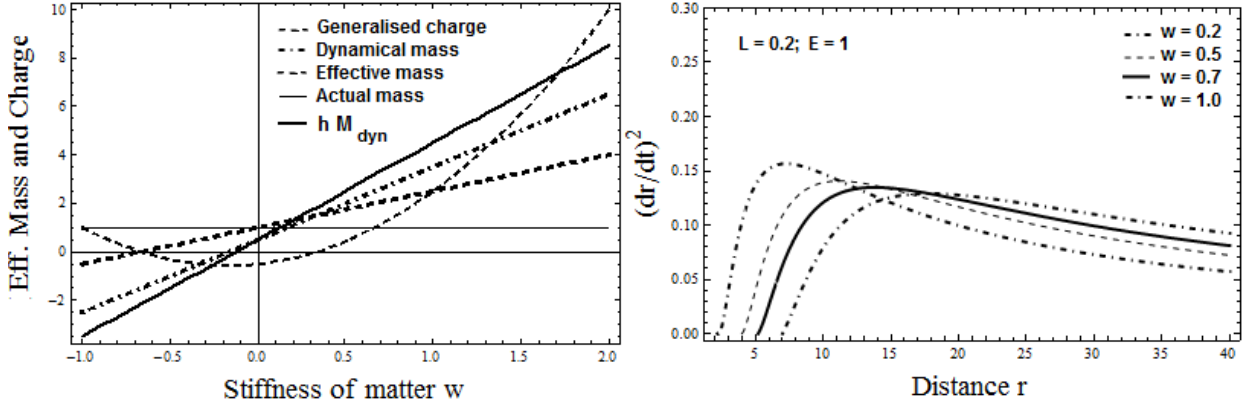}
\caption{(i) Dependence of effective charge and mass parameters on stiffness $w$ (left panel), and (ii) tangential velocity for time-like geodesic case with $M_1G_N=1$ for different values of $w$ (right panel).}
\label{fig7}
 \end{figure}
\noindent The variation of generalised charge, mass parameters and tangential velocity with $w$ is graphically presented in Fig.\ref{fig7}. For non-vanishing values of $A$,  there is a deviation between $M_{dyn}$ and $\tilde{M}$. From the linear analysis and its relation with the general one, for $w=0$, $\tilde{M}$ contributes only half of the dynamical mass (see Fig.\ref{fig7}). $\tilde{M}\approx M_{dyn}$ as measured mass is valid for $w\approx 1/3$ (i.e. $A\approx 0$). The deviation between them enhances with stiffness ($w$), and $\tilde{M}$ grows more rapidly than $M_{dyn}$. The effective masses (i.e. $\tilde{M}$ and $M_{dyn}$) are higher than the actual (density, luminous) mass because of pressure terms themselves. We observe that for higher stiffness of matter (pressure, inner structure), the effective mass $\tilde{M}$ is always higher than $M_{dyn}$. Within linear dynamics, it is the dynamical mass $M_{dyn}$ which dominates, however with non-vanishing values of $w$ according to a PPN framework. The actual effective mass for small pressures ($w$) is approximately given by
\begin{align}
  \tilde{M}= h(w) \, M_{dyn}.
\end{align}
At short distance to gravitational sources and for astrophysical considerations, the dynamical mass gives a measured mass which is unlike the bare, luminous mass from usual density ($M_1$). For $w>1/3$, $\tilde{M}$ is few times higher than the mass $M_1$ or the luminous (density) mass!\\
One may also notice   from Fig.\ref{fig7} a flattening behaviour (though not flat curves) of tangential velocities  ($dr/dt$)  which is as greater as the stiffness. Hence, assuming $R_1M\ll 1$ with $R_1$ as a galaxy radius, higher values of tangential velocities (flattened rotation curves) can be obtained from (\ref{lambda2}) and (\ref{enu2c}) even with simple density profiles (viz Fig.\ref{fig7}). The potency term $B/(2\tilde{M}G_N)$ in (\ref{enu2c}), together with the generalised RN charge $\tilde{Q}^2$ may lead phenomenologically to part of the halo of non-luminous (effective) matter surrounding a galactic core.  Hence, it is reasonable to speculate about a relation to dark-matter phenomenology within this induced model of gravity with Higgs potential. In general, the Cold Dark Matter is believed to generate a linearly growing density distribution so that the tangential velocity remain constant up to the halo radius. The galactic rotation curves in view of a long-range dynamics related to a scalar field within the present model were also investigated recently \cite{Cervantes3,Bezares07-DM}. Flat rotation curves indeed further need to consider more complex density profiles \cite{NFW}, and in fact, for their description one would not only focus on the necessary pressures (which may only contribute on raising rotation curves), but also on the density distribution and scalar-field contributions as Dark Matter in terms of the dark-matter profile including galactic bars \cite{Bezares09-DM, NFW}.

\section{Conclusions and future directions}
In this article, we have investigated the singularities and Black Hole solutions for a model of induced gravity with Higgs potential in view of the Higgs field excitations. We have considered two scenarios (with and without Higgs field mass) to solve the field equations, and the solutions are further analysed in view of the geodesic motion. To conclude, we provide below, in a systematic way, a summary of the results obtained.
\begin{itemize}
\item [(i)] {The Scwarzschild geometry is valid in vacuum for the vanishing scalar field excitations, and all the features of GR are thus applicable for this model of induced gravity in the limit of vanishing excitations.}
\item [(ii)] {We have found the appearance of Reissner--Nordstr\"{o}m-like BH solution for the case of non-vanishing field excitations. There are scalar-field terms (appearing as generalised charge-like parameter) which act anti-gravitationally at low gravitational regimes in RN-like metric.}
\item  [(iii)]  {The terms corresponding to the pressure relevant to the scalar field and nonlinearities of the exact solution lead to a dynamical mass different to the luminous, bare mass derived from the usual  density.}
\item  [(iv)] {The Schwarzschild horizon becomes weaker with stiffness $w$ and the Schwarzschild radius changes with the effective mass. Stiff matter acting repulsively in potential $\lambda$ is an artifact which appears especially for a negative effective, yet positive bare mass.}
\item   [(v)] {The effective potential permits stable bounded
    orbits and angular velocities. The orbits are found qualitatively similar to the case of the Schwarzschild and RN geometry in GR.}
\item  [(vi)] {The assumption of stiff matter (relevant inner
    structure of matter) leads to relevant deviations from effective, measured astronomical masses to bare, luminous masses. It also leads to a raise of curves of tangential velocity.}
\end{itemize}
Still many questions are unanswered from the perspective of the present formulation. Attempts towards analysing solar-relativistic effects and galactic consequences of scalar-field dynamics are of significant importance. The behaviour of flat rotation curves of galaxies and the empiric of Dark Matter (especially for more complex density distributions) have been reported recently \cite{Bezares09-DM}. The primeval dynamics would be another task to investigate within this model, and we intend to report on this issue in our forthcoming communication \cite{nmbup}.

\section*{Acknowledgments}
\small{ One of the authors (HN) is thankful to the University Grants Commission, New Delhi, India for financial support under the UGC--Dr. D. S. Kothari post-doctoral fellowship programme. The work of HN was supported by the German Academic Exchange Service (DAAD), Bonn, Germany in terms of a study visit to the Department of Physics, University of Konstanz, Germany. HN is also thankful
to the HECAP Section, the ASICTP, Italy for the hospitality where a part of this work was initially carried out. He is also grateful to Prof. Anirvan Dasgupta for his suggestion to use series-expansion method to solve the field equations during the early stage of this work. HN and NB would also like to thank Prof. Dr. F. Steiner and the Cosmology and Quantum Gravitation Group of the Institute of Theoretical Physics, as well as the Graduate School for Mathematical Analysis of Evolution, Information and Complexity for their kind support.}
\begin{center}
\noindent\section*{Appendix I}
\end{center}
 \renewcommand{\theequation}{A-\arabic{equation}}
The integration constant $A$ can be obtained by using the scalar-field equation with appropriate boundary conditions. With the equation (\ref{Higgs1}) with vanishing scalar field mass (i.e. $L\rightarrow \infty$) for the asymptotic case (i.e. $r\rightarrow \infty$), we have
\begin{equation}
\nabla_\mu \partial^\mu \xi  = \frac{1}{\sqrt{-g}} \partial_\mu(\sqrt{-g}\, \partial^\mu \xi) = \, \frac{8\pi
\,G}{3}  T;\,\,\,\,\,\,\,\,\,\,\,\,\,\,\,\,\,\,\,\,\,(\hat q=1).\label{a1}
\end{equation}
In the limit $r \to \infty$, $\sqrt{g_{00}} \to 1$ (i.e. a Minkowski spacetime at spatial infinity is valid), with the integration of equation (\ref{a1}) on both the sides, we obtain
\begin{equation}
(\partial_\mu \xi )\, \,4 \pi r^2 = \,-\,\frac{8 \pi G}{3} \, \int T \, \sqrt{-g} \, d^3x .\label{a2}
\end{equation}
For spatial infinity $|e^{(\lambda -\nu )/2}|\sim 1$ and $\partial_\mu \xi \equiv \partial_i \xi= \xi'$ as $\partial_0 \xi = 0$, comparing equations (\ref{H1}) and (\ref{a2}) then leads to
\begin{equation}
A=-\frac{2 G}{3} \, \int T\, \sqrt{-g}\, d^3x. \label{a3}
\end{equation}
\begin{center}
\noindent\section*{Appendix II}
\end{center}
\noindent The scalar-field equation (\ref{Higgs1}) in the limit of vanishing Higgs field  mass reduces to
\begin{equation}
(r^2 e^{(\nu-\lambda)/2}\xi')'=0,  \label{b1}
\end{equation}
which can be written as
\begin{equation}
-\xi''+ \frac{1}{2}(\lambda'- \nu')\xi'= \frac{2}{r}\xi'. \label{b2}
\end{equation}
Adding equations (\ref{F1}) and (\ref{F3}), we obtain
\begin{equation}
F_1 (\xi) + F_3 (\xi) = \frac{1}{1+\xi} \left[-\xi'' + \frac{1}{2}(\lambda'- \nu')\xi'\right]. \label{b3}
\end{equation}
Comparing equations (\ref{b2}) and (\ref{b3}) leads to $F_1 (\xi) +F_3 (\xi)= \frac {2}{r} \frac{\xi'}{1+\xi}$.

\begin{center}
\noindent\section*{Appendix III}
\end{center}
Using the scalar-field equation (\ref{sfes}) and the gravitational potential equation (\ref{E33}) for potential $\Psi$ leads to
\begin{align}
\nabla^2\Psi (1+ \xi)= 4\pi G\left[\epsilon+ 3p- \frac{1}{2}\xi\hat{q} (\epsilon- 3p)\right]- \frac{1}{4}\xi\nabla^2\xi, \label{mp}
\end{align}
In linear approximation, (\ref{mp}) leads to a Poisson equation (\ref{Poisson}). Further, the scalar-field excitation is needed and for a constant energy-density distribution of a homogeneous sphere, the inner-field solution of excitations $\xi_i$ then yields
\begin{equation}
  \xi_i= \frac{1}{r}\left[C_a \sinh(M r)+ C_b \cosh(M r)\right]+ \frac{8\pi G}{3}\frac{\epsilon}{M^2}(1- 3 w);\quad r\leq R_1\,,
\end{equation}
with the integration constants $C_a$ and $C_b$ which as per the continuity conditions at $r=R_1$ are given below,
\begin{equation}
  C_a= -C_b = - \frac{8\pi G}{3 M} (M R_1+ 1) \epsilon (1- 3w)e^{M R_1}.\label{c'c}
\end{equation}
Without a point--mass at the centre, $C_b=0$ and for a gravitating mass--sphere, with $\epsilon=3M_1/(4\pi R_1^3)$, we obtain
\begin{equation}
  C= \frac{2M_1G}{M^3 R_1^2}(1- 3w)\left[M R_1 \cosh(M R_1)- \sinh(M R_1)\right]\,,\label{Cc}
\end{equation}
which for the case $R_1\ll r$ then clearly leads to the value of $C$ as given in equation (\ref{Cconst}).\\
\noindent Now similarly for the potential $\nu$, we have
\begin{align}
  \left(r^2 \nu_i'\right)'+ M^2r^2 \xi_i= \frac{16\pi G}{3}\left(2\epsilon+ 3p\right)r^2.
\end{align}
Integrating the above equation and using the continuity conditions at $r=R_1$ leads to the integration constant $D$ (as given in the equations (\ref{de}) and (\ref{dee})) in the following form,
\begin{equation}
  \hspace{-0.355cm} D= \frac{2M_1 G}{M^2 R^2_1}(1+ 3 w)\left[M^2 R_1^2+ \left(M R_1+ 1\right) \left(1- e^{-M R_1}\right) \left(\cosh(M R_1)- \frac{1}{M R_1}\sinh(M R_1)\right)\right].\label{dcons}
\end{equation}
In fact, with $r\gg R_1$ and general densities, using the scalar field, we obtain
\begin{equation}
 \Phi= -G \int \left(1+ \frac{1}{3}e^{M (|\vec{r}- \vec{r}_S|)}\right) \frac{\epsilon (\vec{r}_S)}{|\vec{r}- \vec{r}_S|}d\vec{r}_S- G \int \left(3- e^{M (|\vec{r}- \vec{r}_S|)} \right) \frac{p(\vec{r}_S)}{|\vec{r}- \vec{r}_S|}d\vec{r}_S\,.
\end{equation}
For point--particles, with $w=\epsilon/p$, for a mass $M_1=4\pi \epsilon R^3$, we have
\begin{equation}
 \Phi=- \frac{M_1 G}{r}\left[1+ 3w+ \frac{1}{3}(1- 3 w)e^{-M r}\right].\label{Phix-an}
\end{equation}
For large length scales, we acquire the relation between the coupling constant $G$ and the Newtonian constant $G_N$ as $G_N=4G/3$ \cite{Dehnen3}. The limit $R_1 M\ll 1$ is valid in torsion experiments for $G_N$.

\end{document}